\documentclass[%
 reprint,
 amsmath,amssymb,
 aps,
prstper,
floatfix,
]{revtex4-2}

\usepackage{graphicx}
\usepackage{dcolumn}
\usepackage{bm}
\usepackage{hyperref}
\usepackage{xcolor}

\begin{document}

\preprint{APS/123-QED}

\title{
Thermalization Rate of Light in Weakly Coupled Molecular Systems
}

\author{Vladislav~Yu.~Shishkov}
\email{vladislav.shishkov@aalto.fi}
\affiliation{
Department of Applied Physics, Aalto University School of Science, FI-00076 Espoo, Finland
}

\date{\today}

\begin{abstract}
Emission and absorption spectra of molecular films are impacted by low-frequency molecular vibrations. 
These vibrations define the linewidths of the absorption and emission spectral peaks, as well as the Stokes shift. 
In cavities that use a molecular film as an active medium, low-frequency molecular vibrations facilitate the thermalization of light, enabling the formation of Bose--Einstein condensation. 
In this work, I employ perturbation theory for Lindblad superoperators and derive the contribution of the low-frequency molecular vibrations to the thermalization rate of light in a weak coupling regime between light and matter.
The derived thermalization rate applies for any cavity design but depends on the local microscopic properties of low-frequency molecular vibrations. 
I provide an estimation for the thermalization rate, which requires only knowledge of the macroscopic parameters of the system: light-matter interaction strength, resonant frequencies of the cavity and excitons, number of molecules in the illuminated area, and the linewidth temperature dependence of the 0-0 peak in the emission spectra of standalone molecular film.
\end{abstract}

\maketitle

\section{Introduction}

Quantum dynamics of active media determine the properties of light in weakly-coupled molecular systems, enabling nonlinearities~\cite{carusotto2013quantum, daskalakis2014nonlinear} that can reach a single-photon level at room temperature~\cite{zasedatelev2021single}.
Different cavity designs, such as planar microcavities with~\cite{schmitt2018dynamics, urbonas2024temporal} or without spatial potential~\cite{putintsev2023controlling, muszynski2024observation}, plasmonic lattices~\cite{hakala2018bose, moilanen2021spatial}, and cavities with bound states in the continuum~\cite{berghuis2023room, yan2025topologically} provide additional flexibility in exploiting these properties of light both in the weak~\cite{schmitt2015thermalization, bloch2022non, dovzhenko2025electrically} and strong coupling regimes~\cite{dovzhenko2018light}.
This broad class of systems enables the realization low-threshold lasers~\cite{kena2010room, ostroverkhova2016organic, keeling2020bose,jiang2022exciton,kavokin2022polariton,luo2023nanophotonics}, optical transistors and logic gates~\cite{zasedatelev2019room, sannikov2024room, de2023room, tassan2024integrated}, and spectrally distant biphoton sources~\cite{shishkov2025entangled}, operating at ambient conditions.

One of the most striking nonlinear properties of light in weakly-coupled molecular systems is thermalization~\cite{rodriguez2013thermalization, kurtscheid2019thermally, vakevainen2020sub, satapathy2022thermalization, peng2023polaritonic}, which facilitates Bose--Einstein condensation~(BEC) of photons and polaritons~\cite{plumhof2014room, cookson2017yellow, busley2022compressibility, dung2017variable, karkihalli2024dimensional, kirton2013nonequilibrium}.
Thermalization of light inside the cavity manifests as a redistribution of the cavity photons over the dispersion surface, conserving the total number of photons~\cite{kavokin2017microcavities}.
The physical picture behind thermalization is the subsequent absorption and re-emittion of light by molecules inside optical cavities~\cite{schmitt2018dynamics}.
The complex dynamics of the molecules between absorption and re-emission events control the thermalization rate of light.
Theoretical analysis of organic molecules reveals that the properties of low-frequency molecular vibrations determine the thermalization rate of light~\cite{kirton2013nonequilibrium, kirton2015thermalization, tereshchenkov2023thermalization}.
These low-frequency molecular vibrations also determine the linewidth of peaks in the emission and absorption spectra and the Stokes shift~\cite{tereshchenkov2023thermalization, karabunarliev2001franck, hoffmann2010determines}.

The formation of polariton (or photon) BEC requires the fulfillment of the following two conditions.
First, the total number of polaritons (or photons) must surpass the critical number, determined for equilibrium BEC~\cite{ketterle1996bose} and recently confirmed for non-equilibrium BEC~\cite{shishkov2022analytical}.
Second, the thermalization of polaritons must be fast enough to form BEC during their lifetime~\cite{shishkov2022exact, shishkov2024sympathetic}.
Thus, a high thermalization rate of light is beneficial because it promotes the formation of BEC.
However, in some cases, a gentle balance between thermalization and loss required to ensure macroscopic occupation in the state of interest.
For instance, some logic gates based on planar microcavities~\cite{zasedatelev2019room, sannikov2024room} require the macroscopic occupation of excited states, i.e. states with ${\bf k} \neq {\bf 0}$.
In this case, high thermalization rate would lead to spontaneous transitions from the excited states to the ground state, preventing the intended operation of these logic gates~\cite{sannikov2024room}.

In the pionering works of thermalization dynamics in plasmonic systems~\cite{kirton2013nonequilibrium, kirton2015thermalization, strashko2018organic}, the joint equations for the occupations of the cavity modes and the saturation of the electronic levels of the molecules allowed to describe the thermalization dynamics of light in plasmonic systems~\cite{hakala2018bose, vakevainen2020sub}.
However, extracting of the thermalization rate in this model requires simulating the system~\cite{hakala2018bose}.
In the recent paper~\cite{tereshchenkov2023thermalization}, we investigated the thermalization rate of polaritons in a strongly coupled molecular systems without spatial potential.
We showed quantitative agreement with the experimental observations but left weakly coupled systems and more complex cavity designs beyond the scope of the work~\cite{tereshchenkov2023thermalization}.

In this paper, I obtain the thermalization rate for weakly-coupled molecular systems with arbitrary cavity designs.
I estimate the thermalization rates relying only on the macroscopic parameters of the system: resonant frequencies of the medium and cavity, light-matter interaction strength, and the temperature dependence of the linewidth of the 0-0 peak in the emission spectrum of standalone molecular films.
For the special case of 2D cavities without spatial potential, the thermalization rate for the light in the weak coupling regime obtained here agrees with the thermalization rate of polaritons in the strong coupling regime obtained in~\cite{tereshchenkov2023thermalization}.


\section{Microscopic theory of molecules inside optical cavities} \label{appendix: weak-coupling}

Electronic and vibrational states hold a central place in the dynamics of molecular systems. 
Despite the energetic disparity between these states, electron-phonon interactions~\cite{hadziioannou2006semiconducting, martinez2020dyes} strongly affect the optical properties of molecules~\cite{gierschner2003optical, bredas2004charge,  reitz2019langevin, chng2024mechanism}. 
When placed inside an optical cavity, electrons, molecular vibrations, and cavity photons can engage in an effective tripartite interaction governed by the microscopic Hamiltonian~\cite{tereshchenkov2023thermalization, shishkov2024sympathetic, shishkov2024room} 
\begin{multline}\label{FullHamiltonian_dressed_approx}
\hat H = 
\sum\limits_{\alpha} 
\hat H_{{\rm cav}{\alpha}}
+
\sum_{m=1}^{N_{\rm mol}} 
\hat H_{\rm exc}^{(m)}
+
\sum_{m=1}^{N_{\rm mol}} 
\sum_{j=1}^{N_{\rm vib}} 
\hat H_{{\rm vib}j}^{(m)}
\\
+
\sum\limits_{m=1}^{N_{\rm mol}} 
\sum\limits_{\alpha} 
\hat H_{{\rm exc}-{\rm cav}{\alpha}}^{(m)}
+
\sum\limits_{m=1}^{N_{\rm mol}} 
\sum_{j=1}^{N_{\rm vib}}
\sum\limits_{\alpha} 
\hat H_{{\rm exc}-{\rm vib}j-{\rm cav}{\alpha}}^{(m)}
\end{multline}
where $N_{\rm mol}$ represents the number of molecules in the system, $N_{\rm vib}$ denotes the number of vibrational modes hosted by each molecule, and index $\alpha$ enumerates the modes of the cavity photons.
The Hamiltonian of the photon mode $\alpha$ is
\begin{equation}
\hat H_{{\rm cav}{\alpha}} = 
\hbar {\omega _{{\rm cav}{\alpha}}}
\hat a_{{\rm cav}{\alpha}}^\dag {{\hat a}_{{\rm cav}{\alpha}}}
\end{equation}
where $\hat a_{{\rm{cav}}{\alpha}}^\dag$ (${\hat a_{{\rm{cav}}{\alpha}}}$) is the creation (annihilation) operator of a photon in the cavity with frequency $\omega_{{\rm{cav}}{\alpha}}$.
These operators obey the bosonic commutation relation ${ \left[ \hat a_{{\rm cav}{\alpha}}, \hat a_{{\rm cav}{\alpha'}}^\dag \right] = \delta_{{\alpha}{\alpha'}} }$.

The Hamiltonian of $m$th molecule hosting a single exciton
\begin{equation}
\hat H_{\rm exc}^{(m)} = 
\hbar \omega_{\rm exc}^{(m)} 
\hat S^{(m)\dag} \hat S^{(m)}
\end{equation}
where the operator $\hat S^{(m)}$ ($\hat S^{(m)\dag}$) is the annihilation (creation) operator of vibrationally dressed excitons with transition energy $\omega_{\rm exc}^{(m)}$, which can be slightly different for each molecule $(m)$ due to the disorder of molecular systems~\cite{bassler1999site, schindler2004universal, hoffmann2010determines}.
The operators of vibrationally dressed excitons are fermionic, obeying the relations $[\hat S^{(m)}, \hat S^{(m')\dag}]=(1-2 \hat S^{(m)\dag} \hat S^{(m)})\delta_{mm'}$ and $\hat S^{(m)}\hat S^{(m)}=0$.

The Hamiltonian of the interaction between vibrationally dressed excitons and cavity photons is
\begin{equation}
\hat H_{{\rm exc}-{\rm cav}{\alpha}}^{(m)} = 
\hbar \Omega_{\alpha}^{(m)}
\left( 
\hat S^{(m)\dag}{{\hat a}_{{\rm cav}{\alpha}}}e^{i \varphi_\alpha^{(m)}}
+
h.c.
\right),
\end{equation}
where $\hbar\Omega_{\alpha}^{(m)} =  - {{\bf E}_{\alpha}^{(m)}}\cdot{\bf d}^{(m)}$ is the single molecule light-matter interaction energy~\cite{scully1997quantum}, with ${\bf E}_{\alpha}^{(m)}$ representing the electric field amplitude for ``one photon'' in the cavity mode $\alpha$ at the position of $m$th molecule, and $\varphi_\alpha^{(m)}$ is the corresponding phase.

I consider a weakly coupled red-detuned organic microcavity, implying $\Omega_{\rm R} < \Gamma_{\rm cav/ 0} < \omega_{\rm exc} - \omega_{\rm cav}$, where $\Omega_{\rm R}$ is the light-matter interaction strength, $\Gamma_{\rm cav}$ is the dissipation rate of the cavity photons, $\Gamma_{\rm 0}$ is the inhomogeneous broadening of excitonic transition in the molecular film, and $\omega_{\rm cav}$ and $\omega_{\rm exc}$ are the natural frequencies of cavity modes and excitons, respectively. 
In this case, the hybrid light-matter exciton-polaritonic states do not form, and I have to consider the cavity photons and excitons separately.

The Hamiltonian of the $j$th mode of excitonically dressed molecular vibrations for the $m$th molecule is
\begin{equation}
\hat H_{{\rm vib}j}^{(m)} = 
\hbar \omega_{{\rm vib}j}^{(m)}\hat B_j^{(m)\dag}\hat B_j^{(m)} 
\end{equation}
where $\omega_{{\rm vib}j}^{(m)}$ is the frequency of the dressed molecular vibration, and $\hat B_j^{(m)}$ ($\hat B_j^{(m)\dag}$) is the corresponding annihilation (creation) operator.
Again, the natural frequencies of the molecular vibrations $\omega_{{\rm vib}j}^{(m)}$ can be slightly different for different molecules due to the disorder of molecular systems~\cite{bassler1999site, schindler2004universal, hoffmann2010determines}.
I order the frequencies of the molecular vibrations $\omega_{{\rm vib}j}^{(m)}$ such that $\omega_{{\rm vib}j}^{(m)} \geq \omega_{{\rm vib}j'}^{(m)}$ for $j>j'$.

The tripartite interaction between dressed excitons localized on the molecule $(m)$, the $j$th dressed molecular vibration of this molecule, and cavity photons in the mode $\alpha$ is~\cite{tereshchenkov2023thermalization, shishkov2024sympathetic, shishkov2024room}
\begin{multline} \label{H exc vib cav}
\hat H_{{\rm exc}-{\rm vib}j-{\rm cav}{\alpha}}^{(m)} = 
\hbar \Lambda_j^{(m)} \Omega _{\alpha}^{(m)}
\left[ 
\hat B_j^{(m)}
\left(
{{\hat a}_{{\rm cav}{\alpha}}}^\dag{e^{-i\varphi_\alpha^{(m)}}}
\hat S^{(m)} 
\right.
\right.
\\
\left.
\left.
-
\hat S^{(m)^\dag}
{{\hat a}_{{\rm cav}{\alpha}}}{e^{i\varphi_\alpha^{(m)}}} 
\right)
+
h.c.
\right]
\end{multline}
where $\Lambda_j^{(m)}$ is the square of Huang--Rhys factor of the $j$th vibrational mode of the $m$th molecule.
It is this tripartite interaction that facilitates the thermalization processes of photons in organic microcavities.

The microscopic theory presented here is valid for any geometry of the cavity, that provides a specific light dispersion $\omega_{{\rm cav}\alpha}$, photonic states $\alpha$, and interaction constants $\Omega_\alpha^{(m)}e^{i\varphi_\alpha^{(m)}}$.
In a specific case of a planar cavity with flat mirrors, I can enumerate photon modes by an in-plane wave vector $\bf k$, using it instead of $\alpha$, and the dispersion relation for the modes around the the ground state would be $\hbar\omega_{{\rm cav}{\bf k}} = \hbar\omega_{{\rm cav}{\bf k}={\bf 0}} + \alpha_{\rm cav} {\bf k}^2$~\cite{keeling2020bose}.
The electric field of the $\bf{k}$th mode, in this case, is distributed in a plane parallel to the mirrors according to $e^{i \bf{kr}}$~\cite{scafirimuto2021tunable, mcghee2021polariton, mcghee2022polariton}; therefore $\varphi_\alpha^{(m)}={\bf k}\cdot{\bf r}^{(m)}$ with ${\bf r}^{(m)}$ pointing to the position of the $m$th molecule.

\section{Separation of the system and reservoir}
In principle, the Hermitian dynamics governed by Hamiltonian~(\ref{FullHamiltonian_dressed_approx}) incorporate the thermalization of light in organic microcavities.
However, the numerous degrees of freedom, including excitons, molecular vibrations, and phonons, makes the direct analysis of the Hermitian dynamics computationally unfeasible. 
To overcome this roadblock and explore the thermalization of the cavity photons, I use the open quantum theory approach~\cite{breuer2002theory}, and I separate the system and reservoir in Hamiltonian~(\ref{FullHamiltonian_dressed_approx}). 

Following~\cite{tereshchenkov2023thermalization}, I separate all the molecular vibrations into two groups: low-frequency molecular vibrations~\cite{colaianni1995low} and high-frequency molecular vibrations.
I denote the upper bound for the frequencies of the low-frequency molecular vibrations as $\omega_{\text{MLFV}}$.
According to the general theory of relaxation in quantum-mechanical systems with closely spaced energy levels, a good choice for this frequency is $\omega_{\text{MLFV}}$ the linewidth of 0-0 transition of the excitons in the molecular film. 
Since I ordered all the vibrational modes such that $\omega_{{\rm vib}n+1}^{(m)} > \omega_{{\rm vib}n}^{(m)}$, I denote the mode $M$ of molecular vibrations that fulfills $\omega_{{\rm vib}M+1}^{(m)} > \omega_{\text{MLFV}} > \omega_{{\rm vib}M}^{(m)}$. 
I identify the Hamiltonian of the reservoir, $\hat H_R$, as the Hamiltonian of the low-frequency molecular vibrations, 
\begin{equation}
\hat H_R = \sum_{m=1}^{N_{\rm mol}} \sum_{j=1}^{M} \hat H_{{\rm vib}j}^{(m)}
.
\end{equation}

I identify the Hamiltonian of the system, $\hat H_S$, as the sum of the Hamiltonian of the cavity photons, $\sum_{\alpha} \hat H_{{\rm cav}{\alpha}}$; excitons, $\sum_{m=1}^{N_{\rm mol}} \hat H_{\rm exc}^{(m)}$; the interaction between cavity photons and excitons, $\sum_{m=1}^{N_{\rm mol}} \sum_{\alpha} \hat H_{{\rm exc}-{\rm cav}{\alpha}}^{(m)}$; high-frequency molecular vibrations, $\sum_{m=1}^{N_{\rm mol}} \sum_{j=M+1}^{N_{\rm vib}}  \hat H_{{\rm vib}j}^{(m)}$; and the interaction between high-frequency molecular vibrations and the excitons, $\sum_{m=1}^{N_{\rm mol}} \sum_{j=M+1}^{N_{\rm vib}} \sum_{\alpha} \hat H_{{\rm exc}-{\rm vib}j-{\rm cav}{\alpha}}^{(m)}$;
\begin{multline} \label{H_S}
\hat H_{S}
=
\sum_{\alpha} \hat H_{{\rm cav}{\alpha}}
+
\sum_{m=1}^{N_{\rm mol}} \hat H_{\rm exc}^{(m)}
+
\sum_{m=1}^{N_{\rm mol}} \sum_{\alpha} \hat H_{{\rm exc}-{\rm cav}{\alpha}}^{(m)}
\\
+
\sum_{m=1}^{N_{\rm mol}} \sum_{j=M+1}^{N_{\rm vib}}  \hat H_{{\rm vib}j}^{(m)}
+
\sum_{m=1}^{N_{\rm mol}} \sum_{j=M+1}^{N_{\rm vib}} \sum_{\alpha} \hat H_{{\rm exc}-{\rm vib}j-{\rm cav}{\alpha}}^{(m)}
.
\end{multline}
The remaining part of Hamiltonian~(\ref{FullHamiltonian_dressed_approx}) is the Hamiltonian of the interaction between the system and the reservoir
\begin{equation} \label{H_SR}
\hat H_{SR} = 
\sum\limits_{m=1}^{N_{\rm mol}} 
\sum_{j=1}^{M}
\sum\limits_{{\alpha}} 
\hat H_{{\rm exc}-{\rm vib}j-{\rm cav}{\alpha}}^{(m)}
\end{equation}
Due to the complexity of the vibrational landscape with its various degrees of freedom, I treat low-frequency vibrations as a reservoir and exclude them using the Born--Markov approximation~\cite{breuer2002theory}. 
This approach implicitly assumes that the heat capacity of the low-frequency molecular vibrations is high enough so that the energy exchange between photons and low-frequency molecular vibrations during the thermalization of photons does not affect the quantum state of the reservoir~\cite{breuer2002theory}, i.e. effective temperature of the low-frequency molecular vibrations remains the same.
A rigorous treatment of the photon thermalization problem should test this assumption on a case-by-case basis.

\section{Thermalization rate of photons}
The identification of the system and reservoir allows us to derive relaxation operators and relaxation rates for the thermalization of photons.
It is evident from Eq.~(\ref{H exc vib cav})~and~(\ref{H_SR}) that there is no direct tripartite interaction between two photons and a molecular vibration. 
Instead, the photons interact with the low-frequency molecular vibrations indirectly through excitons.
Therefore, the {\it local approach} to Lindblad superoperators is unable to describe the thermalization of photons.
To obtain the thermalization rate of photons, I must use the {\it perturbative approach} to Lindblad superoperators instead of the {\it local approach}~\cite{shishkov2020perturbation}.

The perturbation theory developed in~\cite{shishkov2020perturbation} addresses a problem relevant to the thermalization rate analyzed here.
The operators in the system-reservoir interaction Hamiltonian~(\ref{H_SR}), such as $\hat S^{(m)}$, are not eigenoperators.
Therefore, to obtain the correct relaxation operators, I must decompose system operators from~(\ref{H_SR}) into eigenoperators corresponding to the form $\hat A(t) = \sum_{n} \hat A_n e^{-i\omega_n t}$ in the interaction representation~\cite{kosloff2013quantum}. 
For our system, finding the exact decomposition of the operators $\hat S^{(m)}$ into the eigenoperators is a complex problem in itself. 
I overcome this problem by applying {\it perturbative approach} to Lindblad superoperators~\cite{shishkov2020perturbation}, considering ${\Omega_{\alpha}^{(m)}}/{(\omega_{\rm exc}^{(m)}-\omega_{{\rm cav}{\alpha}})}$ as a small parameter.
The approximate Hermitian dynamics of the operators $\hat S^{(m)}$ at low occupation of the excitons, $\langle \hat S^{(m)\dag}\hat S^{(m)} \rangle \ll 1$, can be determined from the Hamiltonian of the system~(\ref{H_S})
\begin{multline}
\hat S^{(m)}(t) =
\hat S^{(m)} e^{-i\omega_{\rm exc}^{(m)}t} 
\\
+ 
\sum_{\alpha} 
\frac{\Omega_{\alpha}^{(m)}e^{i\varphi_\alpha^{(m)}}}{\omega_{\rm exc}^{(m)}-\omega_{{\rm cav}{\alpha}}}
 \hat a_{{\rm cav}{\alpha}} 
\left(
e^{-i\omega_{{\rm cav}{\alpha}}t} - e^{-i\omega_{\rm exc}^{(m)}t}
\right),
\end{multline}
which leads to the approximate time evolution of Hamiltinian~(\ref{H_SR}) in the interaction representation
\begin{multline} \label{H_SR interaction representation}
\hat H_{SR}(t) \approx \hat H_{SR}^{(\rm non-therm)}(t) 
\\
-
\sum\limits_{m=1}^{N_{\rm mol}} 
\sum_{j=1}^{M}
\sum_{\alpha \beta} 
\hbar W^{(m)}_{j{\alpha}{\beta}}
{{\hat a}_{{\rm cav}{\beta}}}^\dag
{{\hat a}_{{\rm cav}{\alpha}}}
e^{-i(\omega_{{\rm cav}{\alpha}}-\omega_{{\rm cav}{\beta}})t}
\\
\left(
\hat B_j^{(m)} e^{-i\omega_{{\rm vib}j}^{(m)}t} 
- 
\hat B_j^{(m)\dag} e^{i\omega_{{\rm vib}j}^{(m)}t}
\right),
\end{multline}
with the effective interaction between photons and low-frequency vibrations,
\begin{equation} \label{W descrete}
W^{(m)}_{j{\alpha}{\beta}} = 
\Lambda_j^{(m)}
\frac
{
(\omega_{{\rm cav}{\alpha}}-\omega_{{\rm cav}{\beta}})
\Omega _{\alpha}^{(m)}\Omega _{\beta}^{(m)}
{e^{i(\varphi_{\beta}^{(m)}-\varphi_{\alpha}^{(m)})}}
}
{(\omega_{\rm exc}^{(m)}-\omega_{{\rm cav}{\beta}})
(\omega_{\rm exc}^{(m)}-\omega_{{\rm cav}{\alpha}})
}.
\end{equation}
In Eq.~(\ref{H_SR interaction representation}), I denote the part of the interaction Hamiltonian that does not contribute to the thermalization processes as $\hat H_{SR}^{(\rm non-therm)}(t)$.
This part of the Hamiltonian contains exciton operators or has a substantial frequency mismatch with the low-frequency vibrations.

Assuming that the low-frequency molecular vibrations are hosted locally on each molecule, $\langle \hat B^{(m)\dag}_j\hat B^{(m')}_{j'} \rangle \propto \delta_{mm'}\delta_{jj'}$, I obtain the Lindblad superoperator $\hat L_{\rm therm}$ for the density matrix of photons, excitons, and high-frequecny molecular vibrations in an organic microcavity, $\hat \rho$~\cite{breuer2002theory, shishkov2020perturbation} 
\begin{multline} \label{Lindbladian}
L_{\rm therm}(\hat \rho) =
\sum_{{\alpha}, {\beta}} 
\frac{\gamma_{\rm therm}^{{\alpha}{\beta}}}{2} 
\left(
2\hat a_{{\rm cav}{\beta }} \hat a_{{\rm cav}{\alpha}}^\dag 
\hat \rho
\hat a_{{\rm cav}{\alpha }} \hat a_{{\rm cav}{\beta }}^\dag
\right.
\\
\left.
-
\hat a_{{\rm cav}{\alpha }} \hat a_{{\rm cav}{\beta }}^\dag
\hat a_{{\rm cav}{\beta }} \hat a_{{\rm cav}{\alpha }}^\dag 
\hat \rho
-
\hat \rho
\hat a_{{\rm cav}{\alpha }} \hat a_{{\rm cav}{\beta }}^\dag
\hat a_{{\rm cav}{\beta }} \hat a_{{\rm cav}{\alpha }}^\dag 
\right),
\end{multline}
where $\gamma_{\rm therm}^{{\alpha}{\beta}}$  is the thermalization rate of the cavity photons, corresponding to the transition from the cavity mode $\beta$ and the cavity mode $\alpha$.
Our theory provides an explicit expression for the thermalization rate
\begin{multline} \label{thermalization rate}
\gamma_{\rm therm}^{\alpha \beta } 
=
2\pi
\sum_{m=1}^{N_{\rm mol}}
\frac
{
\Delta \omega_{\alpha\beta}^2
|\Omega _{\alpha}^{(m)}|^2|\Omega _{\beta}^{(m)}|^2
}
{(\omega_{\rm exc}^{(m)}-\omega_{{\rm cav}{\beta}})^2
(\omega_{\rm exc}^{(m)}-\omega_{{\rm cav}{\alpha}})^2
}
\\
|\Lambda^{(m)}(\Delta \omega_{\alpha\beta})|^2
\nu^{(m)}_{\rm vib}(\Delta \omega_{\alpha\beta})
\left(
1+n_{\rm vib}(\Delta \omega_{\alpha\beta})
\right),
\end{multline}
for $\omega_{{\rm cav} {\beta}}>\omega_{{\rm cav} {\alpha}}$, where $\Delta \omega_{\alpha\beta} = |\omega_{{\rm cav} {\alpha}} - \omega_{{\rm cav} {\beta}} |$, $n_{\rm vib}(\omega) = (e^{\hbar \omega / k_B T} - 1)^{-1}$ is the thermal population of the molecular vibrations with the frequency $\omega$, $k_B$ is the Boltzmann constant, $T$ is the temperature of reservoir and $\nu^{(m)}_{\rm vib}(\omega)$ is the density of states of the vibrational mode with frequency $\omega$ of the molecule $m$. 
I also use the continuous limit of mode distribution of low-frequency vibrations, replacing discrete $\Lambda_j^{(m)}$ in Eq.~(\ref{W descrete}) with continuous $\Lambda^{(m)}(\omega)$.
The ratio of the thermalization rates for the transition from the state $\alpha$ to the state $\beta$ and from the state $\beta$ to the state $\alpha$ follows from the Kubo--Martin--Schwinger relation~\cite{kubo1957statistical} 
\begin{equation} \label{Kubo--Martin--Schwinger}
\gamma_{\rm therm}^{\beta \alpha} 
= 
\gamma_{\rm therm}^{\alpha \beta} 
e^{(\hbar \omega_{\rm cav \alpha} - \hbar \omega_{\rm cav \beta})/k_BT}
.
\end{equation}

The form~(\ref{Lindbladian}) of Lindblad superoperator is well-known and corresponds to the thermalization term in the Maxwell--Boltzmann equation~\cite{yamamoto1999mesoscopic, kavokin2017microcavities}
\begin{multline} \label{thermalization term in MB}
{\rm tr}[L_{\rm therm}(\hat \rho(t)) \hat n_\alpha]
=
\\
\sum_\beta
\left[
\gamma_{\rm therm}^{{\alpha}{\beta}} ( n_\alpha(t) + 1 ) n_\beta(t) 
- 
\gamma_{\rm therm}^{{\beta}{\alpha}} ( n_\beta(t) + 1 ) n_\alpha(t) 
\right]
\end{multline}
where $\hat n_\alpha = \hat a^\dag_{{\rm cav}{\alpha }} \hat a_{{\rm cav}{\alpha }}$, $n_\alpha(t) = {\rm tr} [ \hat n_\alpha \hat\rho(t) ]$, and  ${\rm tr}[...]$ is the trace.
One of the main properties of this Lindblad superoperator is that it conserves the total number of photons, ${\rm tr}[ L_{\rm therm}(\hat \rho) \sum_\alpha  \hat a^\dag_{{\rm cav}{\alpha }} \hat a_{{\rm cav}{\alpha }}]=0$.

A single photon transition from one mode to another due to the thermalization process cannot change the energy of the cavity photon more than by $\hbar \omega_{\text{MLFV}}$ because of the absence of the low-frequency molecular vibrations which could compensate this energy difference.
I account for this by setting $\gamma_{\rm therm}^{\alpha \beta } = 0$ for $\Delta \omega_{\alpha\beta} > \omega_{\text{MLFV}}$ in Eq.~(\ref{thermalization rate}).

Given that the low-frequency molecular vibrations localize on a molecule, the size of the molecules may constrain the possible change in the spatial distribution of the mode during a single transition of photons from one mode to another due to thermalization.
For instance, in planar cavities with flat mirrors, this limits the possible wave vectors of a final cavity mode, ${\bf k}_f$, for a single transition of a photon from an initial state ${\bf k}_i$. 
As an example, I consider MeLPPP molecules, whose size is approximately $l\approx10$~nm~\cite{arkhipov2000ultrafast, hertel1999charge}, implying that the single transition of photons due to thermalization can transfer up to $2\pi/l\approx50~{\rm \mu m^{-1}}$.
This maximum transferred wave vector is larger than the characteristic wave vector of the cavity photons in typical experiments, $\sim 5~{\rm \mu m^{-1}}$~\cite{yoon2022enhanced, plumhof2014room}.
Thus, in this specific case, the size of the molecules barely restricts the thermalization dynamics.

\section{Estimation of thermalization rate}
The application of the Eq.~(\ref{thermalization rate}) to a particular organic microcavity requires knowledge of the local microscopic properties of low-frequency molecular vibrations that facilitate thermalization of photons.
These properties are hard to access, especially in densly packed disordered molecular systems, because the local environment affects these vibrational properties~\cite{boehmke2024uncovering}.
Fortunately, with the theory developed in~\cite{tereshchenkov2023thermalization}, I can address this problem through ensemble-averaged parameters of vibrational modes, excitons, and their coupling to the cavity, and estimate microscopic properties of the molecules in~Eq.~(\ref{thermalization rate}) 
\begin{equation}
\Delta \omega_{\alpha\beta}^2 
|\Lambda^{(m)} (\Delta \omega_{\alpha\beta})|^2 
\nu^{(m)}_{\rm vib}(\Delta \omega_{\alpha\beta})
\approx
\frac{A_2}{\omega_{\text{MLFV}}}
\end{equation}
\begin{equation}
|\Omega _{\alpha}^{(m)}|^2|\Omega _{\beta}^{(m)}|^2
\approx
\frac{|\Omega_{{\rm R}\alpha}|^2|\Omega_{{\rm R}\beta}|^2}{N_{\rm mol}^2}
\end{equation}
where $\Omega_{{\rm R}\alpha}$ is the collective light-matter interaction strength for the mode $\alpha$ (Rabi frequency), and $A_2$ is experimentally accessible parameter of the linewidth of 0-0 emission dependence on the temperature for a standalone molecular film~\cite{tereshchenkov2023thermalization, stampfl1995photoluminescence, guha2003temperature}
\begin{equation}
\Gamma_{\rm exc}^2 \approx \left\{
\begin{array}{lc}
\Gamma_{\rm 0}^2 + C \cdot T, & \; T \gtrsim \hbar \omega_\text{MLFV}/k_B \\
\Gamma_{\rm 0}^2 + A_2, & \; T \ll \hbar \omega_\text{MLFV}/k_B 
\end{array}
\right.
\end{equation}
where $C$ does not depend on the temperature, and $\Gamma_0$ is the inhomogeneous broadening of the emission line.
For instance, for MeLPPP, ${\hbar^2 A_2 \approx 200~{\rm meV^2}}$ and ${\hbar\omega_{\text{MLFV}} \approx 20~{\rm meV}}$~\cite{tereshchenkov2023thermalization}.
I also neglect the difference in exciton transition frequencies~$\omega_{\rm exc}^{(m)}$, replacing them with the mean value~$\omega_{\rm exc}$.
As a result, I obtain the following estimate for the thermalization rate in the case $\omega_{{\rm cav} {\beta}}>\omega_{{\rm cav} {\alpha}}$
\begin{equation} \label{thermalization rate 1}
\gamma_{\rm therm}^{\alpha \beta } 
\approx
\frac
{
2\pi A_2|\Omega_{{\rm R}\alpha}|^2|\Omega_{{\rm R}\beta}|^2
\left(
1+n_{\rm vib}(\omega_{{\rm cav} {\beta}} - \omega_{{\rm cav} {\alpha}})
\right)
}{
N_{\rm mol}
\omega_{\text{MLFV}}
(\omega_{\rm exc}-\omega_{{\rm cav}{\beta}})^2
(\omega_{\rm exc}-\omega_{{\rm cav}{\alpha}})^2
}
\end{equation}
\begin{equation} \label{thermalization rate 2}
\gamma_{\rm therm}^{ \beta \alpha} 
\approx
\frac
{
2\pi A_2|\Omega_{{\rm R}\alpha}|^2|\Omega_{{\rm R}\beta}|^2
n_{\rm vib}(\omega_{{\rm cav} {\beta}} - \omega_{{\rm cav} {\alpha}})
}{
N_{\rm mol}
\omega_{\text{MLFV}}
(\omega_{\rm exc}-\omega_{{\rm cav}{\beta}})^2
(\omega_{\rm exc}-\omega_{{\rm cav}{\alpha}})^2
}
\end{equation}
This result agrees with the estimation obtained in~\cite{tereshchenkov2023thermalization} for a planar organic microcavity with flat mirrors in a strong coupling regime. 

The photons, that ungergo the thermalization, are delocalized quantum states, but the low-frequency molecular vibrations, that facilitate this thermalization process, are localized quantum states. 
Therefore, the spacial overlap of these quantum states, participating in the trepartite interaction, is inverselly proportional to the number of the molecules, defining the $N_{\rm mol}^{-1}$ scale for the thermalization. 

As I discussed in the introduction, one of the conditions for BEC formation, valid both for polaritons and photons, is that the effective thermalization rate must exceed the dissipation rate. 
Thus, the scaling $N_{\rm mol}^{-1}$ of the thermalization rate affects the threshold for BEC formation.
Consider, for instance, a planar organic microcavity with flat mirrors. 
In such organic microcavity, $N_{\rm mol} = n_{\rm mol} \cdot S$, where $n_{\rm mol}$ is the concentration of the molecules, (the number of the molecules per {\it unit area} of the planar cavity), and $S$ is the area of the system, i.e. the area illuminated by external light. 
For a constant concentration of the molecules inside the cavity, the thermalization rate scales as $S^{-1}$, and I should expect the same dependence of the threshold pump {\it power} for BEC formation. 
This is the exact dependence of BEC threshold observed experimentally when the illuminated area is large enough so that the kinetic leakage of the condensate is negligible for its dissipation rate~\cite{putintsev2024photon}.

The detailed analysis of the estimations~(\ref{thermalization rate 1})--(\ref{thermalization rate 2}) shows that they diverge for photon modes with close natural frequencies due to the factor $n_{\rm vib}(\omega_{{\rm cav} {\beta}} - \omega_{{\rm cav} {\alpha}})$.
Specifically, $\gamma_{\rm therm}^{\alpha \beta }$ and $\gamma_{\rm therm}^{\beta \alpha }$ approach infinity as $|\omega_{{\rm cav} {\beta}} - \omega_{{\rm cav} {\alpha}}|$ approaches zero. 
Despite this singularity in the thermalization rate, the macroscopic dynamics of light remain well-defined, as we showed in~\cite{shishkov2025thermalization}, where I analyzed the corresponding Maxwell--Boltzmann equation~\cite{kavokin2017microcavities, deng2010exciton}.

\section{Conclusion and discussion}

In this work, I investigated the microscopic origins of photon thermalization in weakly coupled organic microcavities.
Using the perturbation approach to the Lindblad superoperators~\cite{shishkov2020perturbation}, I derived the thermalization rate of cavity photons due to the complex quantum dynamics of the media affected by the low-frequency molecular vibrations.
The obtained thermalization rate shows agreement with the recent work~\cite{tereshchenkov2023thermalization} in the limit of weak coupling between the excitons and the cavity.
I estimated the thermalization rate by connecting the spectroscopic properties of the organic film and its properties of the low-frequency molecular vibrations.
The developed theory applies not only to 2D organic microcavities with flat mirrors but also to 2D organic microcavities with spatial potentials and to 1D/3D systems.

Generally, there are two main contributions to light thermalization in organic and inorganic microcavities: exciton-exciton scattering and phonon emission~\cite{deng2010exciton, kavokin2017microcavities}.
Organic media host Frenkel excitons, implying the suppression of exciton-exciton scattering processes in organic resonators~\cite{yamamoto2003semiconductor}. 
This fact leaves us with only one contribution to light thermalization in organic cavities: phonon emission.
Thus, I suggest that the thermalization rate obtained here is the dominant contribution to the total thermalization rate of light inside organic microcavities.

\begin{acknowledgments}
Sh.V.Yu. thanks the Magnus Ehrnrooth foundation. 
\end{acknowledgments}

\bibliography{main}

@PREAMBLE{
 "\providecommand{\noopsort}[1]{}" 
 # "\providecommand{\singleletter}[1]{#1}%" 
}

@book{breuer2002theory,
  title={The theory of open quantum systems},
  author={Breuer, Heinz-Peter and Petruccione, Francesco},
  year={2002},
  publisher={Oxford University Press on Demand}
}

@article{reitz2019langevin,
  title={Langevin approach to quantum optics with molecules},
  author={Reitz, Michael and Sommer, Christian and Genes, Claudiu},
  journal={Physical review letters},
  volume={122},
  number={20},
  pages={203602},
  year={2019},
  publisher={APS},
  doi={https://doi.org/10.1103/PhysRevLett.122.203602}
}

@article{schmitt2015thermalization,
  title={Thermalization kinetics of light: From laser dynamics to equilibrium condensation of photons},
  author={Schmitt, Julian and Damm, Tobias and Dung, David and Vewinger, Frank and Klaers, Jan and Weitz, Martin},
  journal={Physical Review A},
  volume={92},
  number={1},
  pages={011602},
  year={2015},
  publisher={APS},
  doi={https://doi.org/10.1103/PhysRevA.92.011602}
}

@book{scully1997quantum,
  title={Quantum optics},
  author={Scully, Marlan O and Zubairy, Suhail},
  year={1997},
  publisher={CambridgeUniversity Press, Cambridge, England}
}

@article{karabunarliev2001franck,
  title={Franck--Condon spectra and electron-libration coupling in para-polyphenyls},
  author={Karabunarliev, Stoyan and Bittner, Eric R and Baumgarten, Martin},
  journal={The Journal of Chemical Physics},
  volume={114},
  number={13},
  pages={5863--5870},
  year={2001},
  publisher={American Institute of Physics},
  doi={https://doi.org/10.1063/1.1351853}
}

@article{zasedatelev2019room,
  title={A room-temperature organic polariton transistor},
  author={Zasedatelev, Anton V and Baranikov, Anton V and Urbonas, Darius and Scafirimuto, Fabio and Scherf, Ullrich and St{\"o}ferle, Thilo and Mahrt, Rainer F and Lagoudakis, Pavlos G},
  journal={Nature Photonics},
  volume={13},
  number={6},
  pages={378--383},
  year={2019},
  publisher={Nature Publishing Group UK London},
  doi={https://doi.org/10.1038/s41566-019-0392-8}
}

@article{plumhof2014room,
  title={Room-temperature Bose--Einstein condensation of cavity exciton--polaritons in a polymer},
  author={Plumhof, Johannes D and St{\"o}ferle, Thilo and Mai, Lijian and Scherf, Ullrich and Mahrt, Rainer F},
  journal={Nature materials},
  volume={13},
  number={3},
  pages={247--252},
  year={2014},
  publisher={Nature Publishing Group UK London},
  doi={https://doi.org/10.1038/nmat3825}
}

@article{zasedatelev2021single,
  title={Single-photon nonlinearity at room temperature},
  author={Zasedatelev, Anton V and Baranikov, Anton V and Sannikov, Denis and Urbonas, Darius and Scafirimuto, Fabio and Shishkov, Vladislav Yu and Andrianov, Evgeny S and Lozovik, Yurii E and Scherf, Ullrich and St{\"o}ferle, Thilo and others},
  journal={Nature},
  volume={597},
  number={7877},
  pages={493--497},
  year={2021},
  publisher={Nature Publishing Group UK London},
  doi={https://doi.org/10.1038/s41586-021-03866-9}
}

@article{scafirimuto2021tunable,
  title={Tunable exciton--polariton condensation in a two-dimensional Lieb lattice at room temperature},
  author={Scafirimuto, Fabio and Urbonas, Darius and Becker, Michael A and Scherf, Ullrich and Mahrt, Rainer F and St{\"o}ferle, Thilo},
  journal={Communications Physics},
  volume={4},
  number={1},
  pages={39},
  year={2021},
  publisher={Nature Publishing Group UK London},
  doi={https://doi.org/10.1038/s42005-021-00548-w}
}

@article{mcghee2021polariton,
  title={Polariton condensation in an organic microcavity utilising a hybrid metal-DBR mirror},
  author={McGhee, Kirsty E and Putintsev, Anton and Jayaprakash, Rahul and Georgiou, Kyriacos and O’Kane, Mary E and Kilbride, Rachel C and Cassella, Elena J and Cavazzini, Marco and Sannikov, Denis A and Lagoudakis, Pavlos G and others},
  journal={Scientific Reports},
  volume={11},
  number={1},
  pages={20879},
  year={2021},
  publisher={Nature Publishing Group UK London},
  doi={https://doi.org/10.1038/s41598-021-00203-y}
}

@article{mcghee2022polariton,
  title={Polariton condensation in a microcavity using a highly-stable molecular dye},
  author={McGhee, Kirsty E and Jayaprakash, Rahul and Georgiou, Kyriacos and Burg, Stephanie L and Lidzey, David G},
  journal={Journal of Materials Chemistry C},
  volume={10},
  number={11},
  pages={4187--4195},
  year={2022},
  publisher={Royal Society of Chemistry},
  doi={https://doi.org/10.1039/D1TC05554B}
}

@article{jiang2022exciton,
  title={Exciton-Polaritons and Their Bose--Einstein Condensates in Organic Semiconductor Microcavities},
  author={Jiang, Zhengjun and Ren, Ang and Yan, Yongli and Yao, Jiannian and Zhao, Yong Sheng},
  journal={Advanced Materials},
  volume={34},
  number={4},
  pages={2106095},
  year={2022},
  publisher={Wiley Online Library},
  doi={https://doi.org/10.1002/adma.202106095}
}

@article{bassler1999site,
  title={Site-selective fluorescence spectroscopy of conjugated polymers and oligomers},
  author={B{\"a}ssler, Heinz and Schweitzer, Bernd},
  journal={Accounts of Chemical Research},
  volume={32},
  number={2},
  pages={173--182},
  year={1999},
  publisher={ACS Publications},
  doi={https://doi.org/10.1021/ar960228k}
}

@book{kavokin2017microcavities,
  title={Microcavities},
  author={Kavokin, Alexey V and Baumberg, Jeremy J and Malpuech, Guillaume and Laussy, Fabrice P},
  volume={21},
  year={2017},
  publisher={Oxford university press}
}

@book{yamamoto2003semiconductor,
  title={Semiconductor cavity quantum electrodynamics},
  author={Yamamoto, Yoshihisa and Tassone, Francesco and Cao, Hui},
  volume={169},
  year={2003},
  publisher={Springer}
}

@article{kirton2015thermalization,
  title={Thermalization and breakdown of thermalization in photon condensates},
  author={Kirton, Peter and Keeling, Jonathan},
  journal={Physical Review A},
  volume={91},
  number={3},
  pages={033826},
  year={2015},
  publisher={APS},
  doi={https://doi.org/10.1103/PhysRevA.91.033826}
}

@article{kirton2013nonequilibrium,
  title={Nonequilibrium model of photon condensation},
  author={Kirton, Peter and Keeling, Jonathan},
  journal={Physical review letters},
  volume={111},
  number={10},
  pages={100404},
  year={2013},
  publisher={APS},
  doi={https://doi.org/10.1103/PhysRevLett.111.100404}
}

@article{yamamoto1999mesoscopic,
  title={Mesoscopic quantum optics},
  author={Yamamoto, Yoshihisa and Imamoglu, Atac},
  journal={Mesoscopic Quantum Optics},
  year={1999}
}

@article{shishkov2022analytical,
  title={Analytical framework for non-equilibrium phase transition to Bose--Einstein condensate},
  author={Shishkov, V Yu and Andrianov, ES and Lozovik, Yu E},
  journal={Quantum},
  volume={6},
  pages={719},
  year={2022},
  publisher={Verein zur F{\"o}rderung des Open Access Publizierens in den Quantenwissenschaften},
  doi={https://doi.org/10.22331/q-2022-05-24-719}
}

@article{shishkov2022exact,
  title={Exact analytical solution for the density matrix of a nonequilibrium polariton Bose-Einstein condensate},
  author={Shishkov, Vladislav Yu and Andrianov, Evgeny S and Zasedatelev, Anton V and Lagoudakis, Pavlos G and Lozovik, Yurii E},
  journal={Physical Review Letters},
  volume={128},
  number={6},
  pages={065301},
  year={2022},
  publisher={APS},
  doi={https://doi.org/10.1103/PhysRevLett.128.065301}
}

@article{kubo1957statistical,
  title={Statistical-mechanical theory of irreversible processes. I. General theory and simple applications to magnetic and conduction problems},
  author={Kubo, Ryogo},
  journal={Journal of the Physical Society of Japan},
  volume={12},
  number={6},
  pages={570--586},
  year={1957},
  publisher={The Physical Society of Japan},
  doi={https://doi.org/10.1143/JPSJ.12.570}
}

@article{guha2003temperature,
  title={Temperature-dependent photoluminescence of organic semiconductors with varying backbone conformation},
  author={Guha, Suchi and Rice, JD and Yau, YT and Martin, Christopher M and Chandrasekhar, Meera and Chandrasekhar, Holalkere R and Guentner, R and De Freitas, P Scanduicci and Scherf, Ullrich},
  journal={Physical Review B},
  volume={67},
  number={12},
  pages={125204},
  year={2003},
  publisher={APS},
  doi={https://doi.org/10.1103/PhysRevB.67.125204}
}

@article{deng2010exciton,
  title={Exciton-polariton bose-einstein condensation},
  author={Deng, Hui and Haug, Hartmut and Yamamoto, Yoshihisa},
  journal={Reviews of Modern Physics},
  volume={82},
  number={2},
  pages={1489},
  year={2010},
  publisher={APS},
  url = {https://doi.org/10.1103/RevModPhys.82.1489}
}

@article{bredas2004charge,
  title={Charge-transfer and energy-transfer processes in $\pi$-conjugated oligomers and polymers: a molecular picture},
  author={Br{\'e}das, Jean-Luc and Beljonne, David and Coropceanu, Veaceslav and Cornil, J{\'e}r{\^o}me},
  journal={Chemical reviews},
  volume={104},
  number={11},
  pages={4971--5004},
  year={2004},
  publisher={ACS Publications},
  doi={https://doi.org/10.1021/cr040084k}
}

@article{gierschner2003optical,
  title={Optical spectra of oligothiophenes: vibronic states, torsional motions, and solvent shifts},
  author={Gierschner, J and Mack, H-G and Egelhaaf, H-J and Schweizer, S and Doser, B and Oelkrug, D},
  journal={Synthetic metals},
  volume={138},
  number={1-2},
  pages={311--315},
  year={2003},
  publisher={Elsevier},
  doi={https://doi.org/10.1016/S0379-6779(03)00030-4}
}

@article{schindler2004universal,
  title={A universal picture of chromophores in $\pi$-conjugated polymers derived from single-molecule spectroscopy},
  author={Schindler, Florian and Lupton, John M and Feldmann, Jochen and Scherf, Ullrich},
  journal={Proceedings of the National Academy of Sciences},
  volume={101},
  number={41},
  pages={14695--14700},
  year={2004},
  publisher={National Acad Sciences},
  doi={https://doi.org/10.1073/pnas.0403325101}
}

@article{hoffmann2010determines,
  title={What determines inhomogeneous broadening of electronic transitions in conjugated polymers?},
  author={Hoffmann, Sebastian T and B{\"a}ssler, Heinz and K{\"o}hler, Anna},
  journal={The Journal of Physical Chemistry B},
  volume={114},
  number={51},
  pages={17037--17048},
  year={2010},
  publisher={ACS Publications},
  doi={https://doi.org/10.1021/jp107357y}
}

@article{stampfl1995photoluminescence,
  title={Photoluminescence and UV-VIS absorption study of poly (para-phenylene)-type ladder-polymers},
  author={Stampfl, J and Graupner, W and Leising, G and Scherf, Ullrich},
  journal={Journal of luminescence},
  volume={63},
  number={3},
  pages={117--123},
  year={1995},
  publisher={Elsevier},
  doi={https://doi.org/10.1016/0022-2313(94)00058-K}
}

@article{shishkov2024sympathetic,
  title={Sympathetic mechanism for vibrational condensation enabled by polariton optomechanical interaction},
  author={Shishkov, Vladislav Yu and Andrianov, Evgeny S and Tretiak, Sergei and Whaley, K Birgitta and Zasedatelev, Anton V},
  journal={Physical review letters},
  volume={133},
  number={18},
  pages={186903},
  year={2024},
  publisher={APS},
  doi={https://doi.org/10.1103/PhysRevLett.133.186903}
}

@article{kosloff2013quantum,
  title={Quantum thermodynamics: A dynamical viewpoint},
  author={Kosloff, Ronnie},
  journal={Entropy},
  volume={15},
  number={6},
  pages={2100--2128},
  year={2013},
  publisher={MDPI},
  doi={https://doi.org/10.3390/e15062100}
}

@article{daskalakis2014nonlinear,
  title={Nonlinear interactions in an organic polariton condensate},
  author={Daskalakis, KS and Maier, SA and Murray, Ray and K{\'e}na-Cohen, St{\'e}phane},
  journal={Nature materials},
  volume={13},
  number={3},
  pages={271--278},
  year={2014},
  publisher={Nature Publishing Group UK London},
  doi = {https://doi.org/10.1038/nmat3874}
}

@article{hakala2018bose,
  title={Bose--Einstein condensation in a plasmonic lattice},
  author={Hakala, Tommi K and Moilanen, Antti J and V{\"a}kev{\"a}inen, Aaro I and Guo, Rui and Martikainen, Jani-Petri and Daskalakis, Konstantinos S and Rekola, Heikki T and Julku, Aleksi and T{\"o}rm{\"a}, P{\"a}ivi},
  journal={Nature Physics},
  volume={14},
  number={7},
  pages={739--744},
  year={2018},
  publisher={Nature Publishing Group},
  doi = {https://doi.org/10.1038/s41567-018-0109-9}
}

@article{moilanen2021spatial,
  title={Spatial and temporal coherence in strongly coupled plasmonic bose-einstein condensates},
  author={Moilanen, Antti J and Daskalakis, Konstantinos S and Taskinen, Jani M and T{\"o}rm{\"a}, P{\"a}ivi},
  journal={Physical Review Letters},
  volume={127},
  number={25},
  pages={255301},
  year={2021},
  publisher={APS},
  doi = {https://doi.org/10.1103/PhysRevLett.127.255301}
}

@article{ostroverkhova2016organic,
  title={Organic optoelectronic materials: mechanisms and applications},
  author={Ostroverkhova, Oksana},
  journal={Chemical reviews},
  volume={116},
  number={22},
  pages={13279--13412},
  year={2016},
  publisher={ACS Publications},
  doi = {https://doi.org/10.1021/acs.chemrev.6b00127}
}

@book{hadziioannou2006semiconducting,
  title={Semiconducting polymers: chemistry, physics and engineering},
  author={Hadziioannou, Georges and Malliaras, George G},
  year={2006},
  publisher={John Wiley \& Sons}
}

@article{rodriguez2013thermalization,
  title={Thermalization and cooling of plasmon-exciton polaritons: towards quantum condensation},
  author={Rodriguez, SRK and Feist, J and Verschuuren, MA and Vidal, FJ Garcia and Rivas, J G{\'o}mez},
  journal={Physical review letters},
  volume={111},
  number={16},
  pages={166802},
  year={2013},
  publisher={APS},
  doi = {10.1103/PhysRevLett.111.166802}
}

@article{satapathy2022thermalization,
  title={Thermalization of Fluorescent Protein Exciton--Polaritons at Room Temperature},
  author={Satapathy, Sitakanta and Liu, Bin and Deshmukh, Prathmesh and Molinaro, Paul M and Dirnberger, Florian and Khatoniar, Mandeep and Koder, Ronald L and Menon, Vinod M},
  journal={Advanced Materials},
  volume={34},
  number={15},
  pages={2109107},
  year={2022},
  publisher={Wiley Online Library},
  doi = {https://doi.org/10.1002/adma.202109107}
}

@article{vakevainen2020sub,
  title={Sub-picosecond thermalization dynamics in condensation of strongly coupled lattice plasmons},
  author={V{\"a}kev{\"a}inen, Aaro I and Moilanen, Antti J and Ne{\v{c}}ada, Marek and Hakala, Tommi K and Daskalakis, Konstantinos S and T{\"o}rm{\"a}, P{\"a}ivi},
  journal={Nature communications},
  volume={11},
  number={1},
  pages={3139},
  year={2020},
  publisher={Nature Publishing Group UK London},
  doi = {https://doi.org/10.1038/s41467-020-16906-1}
}

@article{peng2023polaritonic,
  title={Polaritonic Bottleneck in Colloidal Quantum Dots},
  author={Peng, Kaiyue and Rabani, Eran},
  journal={Nano Letters},
  year={2023},
  publisher={ACS Publications},
  doi = {https://doi.org/10.1021/acs.nanolett.3c03508}
}

@book{martinez2020dyes,
  title={Dyes and photoactive molecules in microporous systems},
  author={Mart{\'\i}nez-Mart{\'\i}nez, Virginia and Arbeloa, Fernando L{\'o}pez},
  year={2020},
  publisher={Springer}
}

@article{shishkov2020perturbation,
  title={Perturbation theory for Lindblad superoperators for interacting open quantum systems},
  author={Shishkov, V Yu and Andrianov, ES and Pukhov, AA and Vinogradov, AP and Lisyansky, AA},
  journal={Physical Review A},
  volume={102},
  number={3},
  pages={032207},
  year={2020},
  publisher={APS},
  doi={https://doi.org/10.1103/PhysRevA.102.032207}
}

@article{tereshchenkov2023thermalization,
url = {https://doi.org/10.1515/nanoph-2023-0800},
title = {Thermalization rate of polaritons in strongly-coupled molecular systems},
title = {},
author = {Evgeny A. Tereshchenkov and Ivan V. Panyukov and Mikhail Misko and Vladislav Y. Shishkov and Evgeny S. Andrianov and Anton V. Zasedatelev},
journal = {Nanophotonics},
doi = {doi:10.1515/nanoph-2023-0800},
year = {2024},
lastchecked = {2024-04-01}
}

@article{keeling2020bose,
  title={Bose--Einstein condensation of exciton-polaritons in organic microcavities},
  author={Keeling, Jonathan and K{\'e}na-Cohen, St{\'e}phane},
  journal={Annual Review of Physical Chemistry},
  volume={71},
  pages={435--459},
  year={2020},
  publisher={Annual Reviews},
  doi={https://doi.org/10.1146/annurev-physchem-010920-102509}
}

@article{strashko2018organic,
  title={Organic polariton lasing and the weak to strong coupling crossover},
  author={Strashko, Artem and Kirton, Peter and Keeling, Jonathan},
  journal={Physical Review Letters},
  volume={121},
  number={19},
  pages={193601},
  year={2018},
  publisher={APS},
  doi={https://doi.org/10.1103/PhysRevLett.121.193601}
}

@article{de2023room,
  title={Room-Temperature Electrical Field-Enhanced Ultrafast Switch in Organic Microcavity Polariton Condensates},
  author={De, Jianbo and Ma, Xuekai and Yin, Fan and Ren, Jiahuan and Yao, Jiannian and Schumacher, Stefan and Liao, Qing and Fu, Hongbing and Malpuech, Guillaume and Solnyshkov, Dmitry},
  journal={Journal of the American Chemical Society},
  volume={145},
  number={3},
  pages={1557--1563},
  year={2023},
  publisher={ACS Publications},
  doi={https://doi.org/10.1021/jacs.2c07557}
}

@article{kavokin2022polariton,
  title={Polariton condensates for classical and quantum computing},
  author={Kavokin, Alexey and Liew, Timothy CH and Schneider, Christian and Lagoudakis, Pavlos G and Klembt, Sebastian and Hoefling, Sven},
  journal={Nature Reviews Physics},
  volume={4},
  number={7},
  pages={435--451},
  year={2022},
  publisher={Nature Publishing Group UK London},
  doi={https://doi.org/10.1038/s42254-022-00447-1}
}

@article{luo2023nanophotonics,
  title={Nanophotonics of microcavity exciton--polaritons},
  author={Luo, Song and Zhou, Hang and Zhang, Long and Chen, Zhanghai},
  journal={Applied Physics Reviews},
  volume={10},
  number={1},
  year={2023},
  publisher={AIP Publishing},
  doi={https://doi.org/10.1063/5.0121316}
}

@article{cookson2017yellow,
  title={A yellow polariton condensate in a dye filled microcavity},
  author={Cookson, Tamsin and Georgiou, Kyriacos and Zasedatelev, Anton and Grant, Richard T and Virgili, Tersilla and Cavazzini, Marco and Galeotti, Francesco and Clark, Caspar and Berloff, Natalia G and Lidzey, David G and others},
  journal={Advanced Optical Materials},
  volume={5},
  number={18},
  pages={1700203},
  year={2017},
  publisher={Wiley Online Library},
  doi={https://doi.org/10.1002/adom.201700203}
}

@article{yoon2022enhanced,
  title={Enhanced thermalization of exciton-polaritons in optically generated potentials},
  author={Yoon, Yoseob and Deschamps, Jude and Steger, Mark and West, Ken W and Pfeiffer, Loren N and Snoke, David W and Nelson, Keith A},
  journal={arXiv preprint arXiv:2209.13703},
  year={2022},
  doi={https://doi.org/10.48550/arXiv.2209.13703}
}

@article{arkhipov2000ultrafast,
  title={Ultrafast on-chain dissociation of hot excitons in conjugated polymers},
  author={Arkhipov, VI and Emelianova, EV and Barth, S and B{\"a}ssler, H},
  journal={Physical Review B},
  volume={61},
  number={12},
  pages={8207},
  year={2000},
  publisher={APS},
  doi={https://doi.org/10.1103/PhysRevB.61.8207}
}

@article{hertel1999charge,
  title={Charge carrier transport in conjugated polymers},
  author={Hertel, D and B{\"a}ssler, H and Scherf, Ullrich and H{\"o}rhold, HH},
  journal={The Journal of chemical physics},
  volume={110},
  number={18},
  pages={9214--9222},
  year={1999},
  publisher={American Institute of Physics},
  doi={https://doi.org/10.1063/1.478844}
}

@article{shishkov2024room,
  title={Room-temperature optomechanics with light-matter condensates},
  author={Shishkov, Vladislav Yu and Andrianov, Evgeny S and Zasedatelev, Anton V},
  journal={Physical Review B},
  volume={110},
  number={13},
  pages={134321},
  year={2024},
  publisher={APS},
  doi={https://doi.org/10.1103/PhysRevB.110.134321}

}

@article{putintsev2023controlling,
  title={Controlling the Spatial Profile and Energy Landscape of Organic Polariton Condensates in Double-Dye Cavities},
  author={Putintsev, Anton D and McGhee, Kirsty E and Sannikov, Denis and Zasedatelev, Anton V and T{\"o}pfer, Julian D and Jessewitsch, Till and Scherf, Ullrich and Lidzey, David G and Lagoudakis, Pavlos G},
  journal={Physical Review Letters},
  volume={131},
  number={18},
  pages={186902},
  year={2023},
  publisher={APS},
  doi={https://doi.org/10.1103/PhysRevLett.131.186902}
}

@article{schmitt2018dynamics,
  title={Dynamics and correlations of a Bose--Einstein condensate of photons},
  author={Schmitt, Julian},
  journal={Journal of Physics B: Atomic, Molecular and Optical Physics},
  volume={51},
  number={17},
  pages={173001},
  year={2018},
  publisher={IOP Publishing},
  doi={10.1088/1361-6455/aad409}
}

@article{carusotto2013quantum,
  title={Quantum fluids of light},
  author={Carusotto, Iacopo and Ciuti, Cristiano},
  journal={Reviews of Modern Physics},
  volume={85},
  number={1},
  pages={299--366},
  year={2013},
  publisher={APS},
  doi={https://doi.org/10.1103/RevModPhys.85.299}
}

@article{sannikov2024room,
  title={Room temperature, cascadable, all-optical polariton universal gates},
  author={Denis A. Sannikov and Anton V. Baranikov and Anton D. Putintsev and Mikhail Misko and Anton V. Zasedatelev and Ullrich Scherf and Pavlos G. Lagoudakis},
  journal={Nature communications},
  volume={15},
  number={5},
  pages={5362},
  year={2024},
  publisher={Nature Publishing Group UK London},
  doi={https://doi.org/10.1038/s41467-024-49690-3}
}

@article{tassan2024integrated,
  title={Integrated ultrafast all-optical polariton transistors},
  author={Tassan, Pietro and Urbonas, Darius and Chmielak, Bartos and Bolten, Jens and Wahlbrink, Thorsten and Lemme, Max C and Forster, Michael and Scherf, Ullrich and Mahrt, Rainer F and St{\"o}ferle, Thilo},
  journal={arXiv preprint arXiv:2404.01868},
  year={2024},
  doi={https://doi.org/10.48550/arXiv.2404.01868}
}

@article{muszynski2024observation,
  title={Observation of a stripe phase in a spin-orbit coupled exciton-polariton Bose-Einstein condensate},
  author={Muszynski, Marcin and Kokhanchik, Pavel and Urbonas, Darius and Kapuscinski, Piotr and Oliwa, Przemyslaw and Mirek, Rafal and Georgakilas, Ioannis and Stoferle, Thilo and Mahrt, Rainer F and Forster, Michael and others},
  journal={arXiv preprint arXiv:2407.02406},
  year={2024},
  doi={https://doi.org/10.48550/arXiv.2407.02406}
}

@article{urbonas2024temporal,
  title={Temporal mode switching during polariton condensation},
  author={Urbonas, Darius and Moilanen, Antti J and Arnardottir, Kristin B and Scherf, Ullrich and Mahrt, Rainer F and T{\"o}rm{\"a}, P{\"a}ivi and St{\"o}ferle, Thilo},
  journal={Communications Physics},
  volume={7},
  number={1},
  pages={203},
  year={2024},
  publisher={Nature Publishing Group UK London},
  doi={https://doi.org/10.1038/s42005-024-01667-w}
}

@article{shishkov2025thermalization,
  title={Thermalization in Quantum Fluids of Light: A Convection-Diffusion Equation},
  author={Shishkov, Vladislav Yu and Panyukov, Ivan V and Andrianov, Evgeny S and Zasedatelev, Anton V},
  journal={arXiv preprint arXiv:2501.10537},
  year={2025},
  doi={https://doi.org/10.48550/arXiv.2501.10537}
}

@article{shishkov2025entangled,
  title={Entangled Polariton States in the Visible and Mid-Infrared Spectral Ranges},
  author={Shishkov, Vladislav Yu and Kotov, Oleg and Haughton, Emily and Urbonas, Darius and Rozema, Lee A and Garcia-Vidal, Francisco J and Feist, Johannes and Zasedatelev, Anton V},
  journal={arXiv preprint arXiv:2508.10809},
  year={2025},
  doi={https://doi.org/10.48550/arXiv.2508.10809}
}

@article{dovzhenko2025electrically,
  title={Electrically reconfigurable extended lasing state in an organic liquid-crystal microcavity},
  author={Dovzhenko, Dmitriy and Ricco, Luciano Siliano and Sawicki, Krzysztof and Muszy{\'n}ski, Marcin and Kokhanchik, Pavel and Kapu{\'s}ci{\'n}ski, Piotr and Morawiak, Przemys{\l}aw and Piecek, Wiktor and Nyga, Piotr and Kula, Przemys{\l}aw and others},
  journal={arXiv preprint arXiv:2506.05717},
  year={2025},
  doi={https://doi.org/10.48550/arXiv.2506.05717}
}

@article{dovzhenko2018light,
  title={Light--matter interaction in the strong coupling regime: configurations, conditions, and applications},
  author={Dovzhenko, DS and Ryabchuk, SV and Rakovich, Yu P and Nabiev, IR},
  journal={Nanoscale},
  volume={10},
  number={8},
  pages={3589--3605},
  year={2018},
  publisher={Royal Society of Chemistry},
  doi={https://doi.org/10.1039/C7NR06917K}
}

@article{kena2010room,
  title={Room-temperature polariton lasing in an organic single-crystal microcavity},
  author={K{\'e}na-Cohen, St{\'e}phane and Forrest, SR},
  journal={Nature Photonics},
  volume={4},
  number={6},
  pages={371--375},
  year={2010},
  publisher={Nature Publishing Group UK London},
  doi={https://doi.org/10.1038/nphoton.2010.86}
}

@article{bloch2022non,
  title={Non-equilibrium Bose--Einstein condensation in photonic systems},
  author={Bloch, Jacqueline and Carusotto, Iacopo and Wouters, Michiel},
  journal={Nature Reviews Physics},
  volume={4},
  number={7},
  pages={470--488},
  year={2022},
  publisher={Nature Publishing Group UK London},
  doi={https://doi.org/10.1038/s42254-022-00464-0}
}

@article{busley2022compressibility,
  title={Compressibility and the equation of state of an optical quantum gas in a box},
  author={Busley, Erik and Miranda, Leon Espert and Redmann, Andreas and Kurtscheid, Christian and Umesh, Kirankumar Karkihalli and Vewinger, Frank and Weitz, Martin and Schmitt, Julian},
  journal={Science},
  volume={375},
  number={6587},
  pages={1403--1406},
  year={2022},
  publisher={American Association for the Advancement of Science},
  doi={10.1126/science.abm2543}
}

@article{dung2017variable,
  title={Variable potentials for thermalized light and coupled condensates},
  author={Dung, David and Kurtscheid, Christian and Damm, Tobias and Schmitt, Julian and Vewinger, Frank and Weitz, Martin and Klaers, Jan},
  journal={Nature Photonics},
  volume={11},
  number={9},
  pages={565--569},
  year={2017},
  publisher={Nature Publishing Group UK London},
  doi={https://doi.org/10.1038/nphoton.2017.139}

}

@article{karkihalli2024dimensional,
  title={Dimensional crossover in a quantum gas of light},
  author={Karkihalli Umesh, Kirankumar and Schulz, Julian and Schmitt, Julian and Weitz, Martin and von Freymann, Georg and Vewinger, Frank},
  journal={Nature Physics},
  volume={20},
  number={11},
  pages={1810--1815},
  year={2024},
  publisher={Nature Publishing Group UK London},
  doi={https://doi.org/10.1038/s41567-024-02641-7}
}

@article{kurtscheid2019thermally,
  title={Thermally condensing photons into a coherently split state of light},
  author={Kurtscheid, Christian and Dung, David and Busley, Erik and Vewinger, Frank and Rosch, Achim and Weitz, Martin},
  journal={Science},
  volume={366},
  number={6467},
  pages={894--897},
  year={2019},
  publisher={American Association for the Advancement of Science},
  doi={https://doi.org/10.1126/science.aay1334}
}

@article{berghuis2023room,
  title={Room temperature exciton--polariton condensation in silicon metasurfaces emerging from bound states in the continuum},
  author={Berghuis, Anton Matthijs and Castellanos, Gabriel W and Murai, Shunsuke and Pura, Jose Luis and Abujetas, Diego R and van Heijst, Erik and Ramezani, Mohammad and S{\'a}nchez-Gil, Jos{\'e} A and Rivas, Jaime G{\'o}mez},
  journal={Nano letters},
  volume={23},
  number={12},
  pages={5603--5609},
  year={2023},
  publisher={ACS Publications},
  doi={https://doi.org/10.1021/acs.nanolett.3c01102}
}

@article{yan2025topologically,
  title={Topologically reconfigurable room-temperature polariton condensates from bound states in the continuum in organic metasurfaces},
  author={Yan, Xingchen and Tang, Min and Zhou, Zhonghao and Ma, Libo and Vaynzof, Yana and Yao, Jiannian and Dong, Haiyun and Zhao, Yong Sheng},
  journal={Nature Communications},
  volume={16},
  number={1},
  pages={1--8},
  year={2025},
  publisher={Nature Publishing Group},
  doi={https://doi.org/10.1038/s41467-025-57738-1}
}

@article{colaianni1995low,
  title={Low-frequency Raman spectroscopy},
  author={Colaianni, SE May and Nielsen, O Faurskov},
  journal={Journal of molecular structure},
  volume={347},
  pages={267--283},
  year={1995},
  publisher={Elsevier},
  doi={https://doi.org/10.1016/0022-2860(95)08550-F}
}

@article{boehmke2024uncovering,
  title={Uncovering low-frequency vibrations in surface-enhanced Raman of organic molecules},
  author={Boehmke Amoruso, Alexandra and Boto, Roberto A and Elliot, Eoin and de Nijs, Bart and Esteban, Ruben and F{\"o}ldes, Tam{\'a}s and Aguilar-Galindo, Fernando and Rosta, Edina and Aizpurua, Javier and Baumberg, Jeremy J},
  journal={Nature Communications},
  volume={15},
  number={1},
  pages={6733},
  year={2024},
  publisher={Nature Publishing Group UK London},
  doi={https://doi.org/10.1038/s41467-024-50823-x}
}

@article{putintsev2024photon,
  title={Photon statistics of organic polariton condensates},
  author={Putintsev, Anton D and Zasedatelev, Anton V and Shishkov, Vladislav Yu and Misko, Mikhail and Sannikov, Denis A and Andrianov, Evgeny S and Lozovik, Yurii E and Scherf, Ullrich and Lagoudakis, Pavlos G},
  journal={Physical Review B},
  volume={110},
  number={4},
  pages={045125},
  year={2024},
  publisher={APS},
  doi={https://doi.org/10.1103/PhysRevB.110.045125}
}

@article{chng2024mechanism,
  title={Mechanism of molecular polariton decoherence in the collective light--matter couplings regime},
  author={Chng, Benjamin XK and Ying, Wenxiang and Lai, Yifan and Vamivakas, A Nickolas and Cundiff, Steven T and Krauss, Todd D and Huo, Pengfei},
  journal={The Journal of Physical Chemistry Letters},
  volume={15},
  number={47},
  pages={11773--11783},
  year={2024},
  publisher={ACS Publications},
  doi={https://doi.org/10.1021/acsphotonics.7b00728}
}

@article{ketterle1996bose,
  title={Bose-Einstein condensation of a finite number of particles trapped in one or three dimensions},
  author={Ketterle, Wolfgang and Van Druten, NJ},
  journal={Physical Review A},
  volume={54},
  number={1},
  pages={656},
  year={1996},
  publisher={APS},
  doi={https://doi.org/10.1103/PhysRevA.54.656}
}

\end{document}